# Towards pristine graphene-metal interface and microstructures: Laser assisted direct patterning on Epitaxial graphene

A. Nath*[1], M. Currie[2], V.D. Wheeler[2], M.J. Tadjer[2], A.D. Koehler[2], Z.R. Robinson[2], K. Sridhara[3], S. C. Hernandez[2], J. A. Wollmershauser[2], J. T Robinson[2], R.L. Myers-Ward[2], C.R. Eddy, Jr.[2], M.V.Rao[1] and D.K. Gaskill[2]

[1]*George Mason University, 4400 University Dr. Fairfax, Virginia, VA 22030, USA*

[2]*U.S. Naval Research Laboratory, 4555 Overlook Ave., SW, Washington, DC 20375, USA*

[3]*University of Maryland, College Park, MD 20742, USA*

**Abstract:** Graphene-metal contact resistance is governed by both intrinsic and extrinsic factors. Intrinsically, both the density of states bottleneck near the Dirac point and carrier reflection at the graphene-metal interface lead to a high contact resistance. Moreover, graphene exhibits insulating behavior for out-of-the-plane conduction. Extrinsically, surface contamination introduced by photoresist residue or different adsorbed species during standard lithography processing alters graphene's intrinsic properties by uncontrolled doping and increased scattering which results in high and inconsistent contact resistance. Here we demonstrate a femto-second laser assisted direct patterning of graphene microstructures that enables us to study both intrinsic and extrinsic effects on the graphene-metal interface. We show that a clean graphene-metal interface is not sufficient to obtain contact resistance approaching the intrinsic limit set by the quantum resistance. We also demonstrated that unlike CVD graphene, edge state conduction (or end-contact) is not spontaneously formed by metal deposition in case of graphene grown on SiC(0001). We conclude that for epitaxial graphene, intentional end-contact formation is necessary to obtain contact resistance near the quantum contact resistance limit.

*Corresponding author's email: anath@gmu.edu

Understanding and controlling the graphene-metal interaction poses an intriguing challenge for the graphene community. Controlled tailoring of the graphene-metal contact resistance ($R_C$) is essential for applications that exploit its extraordinary electronic, optical, thermal and mechanical properties.[1,2] Both intrinsic and extrinsic effects contribute to graphene-metal $R_C$. Intrinsically, even though graphene is a semi-metal, the density of states (DOS) bottleneck near the Dirac point leads to an elevated $R_C$.[3] Moreover it has been suggested that the successive transformation between Dirac-like and Schrodinger-like carriers at the graphene-metal interface decreases the carrier transmission probability which results in a higher $R_C$.[4] Extrinsically, surface contamination introduced by polymers, solvents, chemicals and other adsorbates such as water-vapor during standard semiconductor processing modify the intrinsic properties of graphene through increased scattering.[5] Resist residue at the interface between the graphene and the metal has been shown to inhibit conformal deposition of metal on graphene, resulting in increased and inconsistent $R_C$.[6] Yet, unlike conventional semiconductors, resist residue on graphene surface cannot be cleaned using standard plasma-ashing due to the low selectivity between carbon-based resists and graphene.[7] Also, p-n junction formation due to different work function between graphene and metal contacts may contribute significantly to $R_C$.[8]

In recent years, various strategies have been employed to achieve reproducible low contact resistance. Several groups have attributed low $R_C$ to a clean graphene-metal interface.[9,10] Other approaches have enhanced the interfacial DOS and/or graphene-metal carrier transmission by intentionally damaging graphene by oxygen plasma,[11] ultraviolet/ozone treatment,[12] contact area patterning,[13] or by one-dimensional side contact.[14] Additionally, Wallace *et al*, used *in-situ* X-ray photoelectron spectroscopy (XPS) to show that metal deposition (Ti and Pd) on as-grown chemical vapor deposition (CVD) graphene spontaneously form defects resulting in end-contact (i.e. conduction through reactive edge-states or defects). [15] Another viable approach to tailor $R_C$ is metal-graphene work function engineering.[16] Several groups have achieved relatively low contact resistance using Pd or Ni and higher contact resistance with Ti, Cr, and Al contacts.[17] Yet, Ti/Au contact resistances approaching quantum $R_C$ values were recently reported. [7,9] It has been theoretically suggested that a change in the graphene dispersion relation occurs when the metal chemisorbs on the graphene (Ti, Ni, Co, Cr and Pd) and not when the metal physisorbs (Au, Ag, and Pt) [3,15] contradicting other experimental findings. [17] In contrast, Robinson et al.[11] have reported no significant dependence of $R_C$ with contact metal.

One of the major limitations of prior contact resistance reports is that the graphene was always contaminated and/or modified before metal deposition by lithography resists, plasma damage, etc. Recently two different groups [18,19] have reported a resist-free process to study residue free graphene-metal interface. Yet, in both studies, exfoliated graphene was used and due to the small size of the flakes end-contacting of the graphene was unavoidable.[19] In this work, we have developed a resist-free technique, which when used on large area epitaxial graphene (EG) grown on SiC permits the study of graphene–metal interaction without process induced artifacts such as resist residue or end-contacts due to mesa formation. Using this approach, we are able to determine whether a resist-free interface is sufficient and/or necessary to obtain a low $R_C$ (limited by quantum contact resistance), and to verify if spontaneous defect formation during Ni metal deposition is inherent to graphene.

The graphene samples were prepared by silicon sublimation from a semi-insulating SiC (0001) substrate in an Ar atmosphere. The growth conditions have previously been shown to result in uniform graphene thickness on terraces where the bounding steps contain an extra layer.[20]

Van der Pauw (vdP) Hall and transfer length measurement (TLM) structures were fabricated by a photoresist-free process implemented with a custom made shadow mask (Fig. 1(a)). In order to fabricate the shadow mask for contact metal deposition, a one µm thick $SiO_2$ film was deposited by plasma enhanced CVD (Oxford Instruments PECVD) on a standard 4-inch Si (100) wafer thinned to 100 µm by chemical-mechanical polishing. The test structures were then patterned on to the $SiO_2$/Si wafer by standard photolithography. The exposed $SiO_2$ was subsequently etched by a combination of wet-etch (buffered oxide etch) and dry-etch in a commercial deep reactive ion etching chamber (Oxford Instruments DRIE, 1000 W ICP, 150 W RF, 50 sccm $CHF_3$, 30 sccm $O_2$, 40 mT, 20°C). An $SF_6$-based chemistry was used to cryo-etch through the silicon wafer anisotropically in the same DRIE chamber (1000 W ICP, 9 W RIE, 100 sccm $SF_6$, 5 sccm $O_2$, 15 mT, -110 °C).[21] The resulting $SiO_2$/Si mask was then placed on the graphene surface and a 50 nm thick Ni film was deposited by electron-beam evaporation (Fig. 1(b)).

For device isolation (Fig. 1(c)), the samples were illuminated by optical pulses from an amplified Ti:sapphire laser (Coherent RegA) at a repetition rate of 250 kHz with 800-nm center wavelength and 35-nm optical bandwidth (full width at half maximum).[22] An optical pulse shaper located between the Ti:sapphire seed laser and amplifier pre-compensated for the system dispersion. BioPhotonic Solution's MIIPS system implemented the necessary feedback to the pulse shaper to produce 50-fs pulses at the sample.

The laser was focused with a 50x objective lens to approximately 1-µm-diameter focal spot. Due to the tight focus, the Rayleigh range is also on the order of 1 µm. Thus, to maintain the focal spot while laterally scanning, the sample requires <1 µm of height variation between our laser and the sample. To accomplish this, the sample was positioned on a 5-axis stage. The laser writing was performed by opening an electro-mechanical shutter and moving the sample in the two lateral dimensions to produce the desired pattern. Ablation of the graphene on a sub-micrometer scale was achieved with an average optical irradiance < 80 kW cm$^{-2}$, more than 4x less than for damage (not ablation) induced in graphene by continuous wave (cw) lasers. [23] Our average optical irradiance corresponds to < 3 nJ pulse energy and < 0.3 J cm$^{-2}$ optical pulse fluence.

Post isolation device electrical measurements were performed in a custom low vacuum (10$^{-4}$ mbar) probe station. Measurements were performed *in-situ* at room temperature (RT) before and after the sample was subjected to a 24 hr vacuum anneal at 200 °C. Immediately after the *in-situ* anneal, RT Hall measurements were performed in air on $12 \times 12$ μm$^2$ vdP structures adjacent to the TLM structures. Post annealing characterization included scanning electron microscopy (SEM, Carl Zeis) and Raman spectroscopy (Thermo DXR) using a 9 mW 532 nm laser. Kelvin probe force microscopy (KPFM) was performed on a Veeco D5000 scanning probe microscope in ambient conditions using a double pass technique with Co/Cr-coated silicon tip (radius of curvature of ~50 nm). Atomic force microscopy (AFM, Bruker Dimension Icon) and optical microscopy (Olympus BX51, edge detection Sobel filter mode) were employed to measure TLM pad distances (150 μm width, spacing range was 5-60 μm).

Fig. 2(a) depicts a SEM image of an isolated TLM structure before *in-situ* anneal. The deposited metal was conformal and extremely smooth with rms roughness ≈ 0.25 nm (Fig. 2(a)) inset). The isolation line width was about 800 nm as measured by SEM (not shown) and AFM (Fig. 2(b)). A Raman map of the graphene 2D peak intensity (Fig. 2(c)) shows the complete removal of graphene from the areas damaged with the laser. Some debris were produced during laser isolation and could be observed in the vicinity (≈1μm) of the isolation line, however, the active regions of the devices were free from debris, as can be seen in Fig. 2(d). It should be noted that the graphene immediately neighboring the isolation line was not altered due to the use of a pulsed laser. Here, graphene is ablated on a femtosecond time scale where the irradiation is 4x lower than needed to damage the graphene with cw laser (the cw laser damage is likely due to local heating).[23] For cw lasers, the high heat capacity and thermal conductivity of graphene influence the ablation process and create a damaged area much larger than the laser spot size. However, pulsed lasers have the potential for nonlinear absorption. Their subpicosecond energy absorption times are much faster than thermal or acoustic processes, thereby enabling patterning that has smaller damage areas with sharper boundaries.

This nonlinear optical effect depends on the irradiance/fluence of the optical beam, thus, the size of the damaged region is controlled by the laser's spatial profile. A common metric for focused Gaussian beams is the diameter measured at the points where the irradiance drops to $1/e^2$ of the peak value. However, a peak irradiance (e.g., >80% of the maximum value) occurs over a smaller diameter. Our 1-μm diameter laser spot (measured at $1/e^2$) used in this study produced a somewhat smaller damaged region in the graphene film. This is explained by our nonlinear damage and laser-ablated region occurring at an irradiance greater than at the focussed laser's 1-μm ($1/e^2$) diameter, thereby producing a damaged region smaller than the $1/e^2$ diameter.

Fig. 3(a) shows an optical image of a representative Hall structure. The use of thinned Si wafer and optimized cryo-etch facilitated a shadow mask with a resolution ≈ 2 μm and SEM images (not shown) revealed no ragged edges indicating smooth sidewalls of the deposited metal. The Raman 2D peak full width at half maximum (FWHM) map of an isolated vdP cross is shown in Fig. 3(b), where the isolation line is evident due to the absence of graphene 2D peaks. Fig. 3(c) depicts AFM height image of a TLM structure. In this AFM image, a shadowing effect was observed due to the finite gap between the physical mask and graphene. However, it was ≈ 1 μm and included in the TLM calculation.

The KPFM image of the same TLM structure (Fig. 3(c)) reveals an average work function difference, $\Delta\phi = \phi_{Ni} - \phi_{graphene} \approx 0.35$ eV between Ni pads and graphene. This difference is roughly consistent with prior measurements of the Ni-(CVD) graphene work function difference of -0.2 to 0.4 eV by ultraviolet photoelectron spectroscopy (UPS). [16]

Fig. 4(a) depicts the RT I-V characteristics for various TLM separations of a representative device after annealing for 24 hr at 200° C. Ohmic behavior was observed for all three measured devices both before and after annealing. The inset of the Fig. 4(a) shows the I-V behavior between two adjacent TLM structures where the resistance between two adjacent but isolated devices is >$10^{12}$ Ω, which indicates good isolation. Fig 4. (b) plots the total resistance as a function of TLM pad spacing before and after *in-situ* annealing. The contact resistances were calculated to be 2625± 105 Ω·μm and 1200 ± 107 Ω·μm before and after annealing, respectively. However, the sheet resistance under the metal before and after annealing were nearly identical, ≈ 840 Ω/□ as found from the slope of the TLM measurements.

The conductance (G) of a graphene-metal junction can be described by the Landauer-Buttiker model for one dimensional wire, $G = \frac{2e^2}{h}TM$ [24,25] where $T$ is the carrier transmission probability, M is the conduction mode in graphene, $e$ is the electron charge and $h$ is Planck's constant. Considering two valleys of graphene, it can be shown that the graphene-metal contact resistance is, $R_C = \frac{1}{T}\frac{h\pi^{\frac{1}{2}}}{4e^2 n^{\frac{1}{2}}}$, [14,24]     (1) where $n$ is the sheet density underneath the metal. We can calculate the quantum limited contact resistance from Eq. (1) by assuming perfect transmission ($T=1$) and using the measured Hall carrier concentration as an approximation of $n$. The room temperature $12 \times 12$ μm$^2$ vdP measurement showed an electron concentration $\approx 1.7 \times 10^{12}$ cm$^{-2}$, mobility $\approx 1700$ cm$^2$ V$^{-1}$.s$^{-1}$ and sheet resistance ($R_{Sh}$) $\approx 2000$ Ω/□. This results in a $R_C$ of 88 Ω-μm, which is 12.5x lower than the value we obtained after annealing (1200 ± 107 Ω·μm).

The origin of such disparity can be attributed to reduction in carrier transmission probability for two reasons, neither of which is accounted for in our calculation. The carrier transmission probability ($T$) depends on $T_{M-G}$ (transport from the metal into the graphene) and $T_C$ (transport from graphene beneath the metal to the graphene channel), which can be expressed as $T = T_C T_{M-G}/(1-(1-T_C)(1-T_{M-G}))$.[25] The first reason for reduction is, even though graphene is a semi-metal and exhibits high lateral conductivity, it essentially serves as an insulator for out-of-the plane conduction.[26] Hence, in the absence of edge-state conduction, the $T_{M-G}$ is reduced significantly. The second reason for reduction stems from the possible formation of p-n junction at the interface of graphene underneath the metal and the bare graphene. Theoretical work predicts that metal induced graphene doping depends on both the work function difference and graphene-metal surface distance.[27] In this study, the graphene Fermi level ($\varepsilon_F$) resides 0.35 eV above the Ni Fermi level, and it is expected that the graphene underneath the metal is p doped because of charge transfer, as recently shown experimentally by Yang *et al.*[16] This effect is also manifested by our RT vdP Hall measurement showing the sheet resistances under the Ni were markedly different than the bare graphene. Since the graphene channel remains n-type, this will create a p-n junction. The transmission probability ($T_C$) through such a p-n junction will be smaller than unity.[25] Moreover intrinsic effects such as the momentum mismatch between the metal and graphene, finite graphene-metal distance, successive transformation from Dirac-like to Schrodinger-type carriers at the interface, and carrier reflection due to non-normal incident angles also impede carrier transmission and should result in a higher contact resistance.

In this study, a significant improvement (55%) of the $R_C$ was achieved by vacuum annealing. From Eq. 1, contact resistance can be improved either by elevated carrier concentration or by enhanced carrier transmission. Since $R_{Sh}$ is primarily dictated by the carrier concentration and vacuum anneal reduces adsorbates (which are responsible for weak p-type doping in graphene[28]) at the graphene-metal interface, annealing generally results in a reduction of $R_{Sh}$ underneath the metal.[7] This phenomenon was not observed in this study. We postulate that $R_{Sh}$ (hence *n*) was not changed by annealing either due to the absence of any extrinsic adsorbates at the interface or because of graphene Fermi level pinning due to its strong interaction with Ni. [17] Hence we attribute the improvement of $R_C$ to an enhancement of the carrier transmission, which is consistent with the previous studies.[13] The enhancement in transmission is likely due to improved proximity of the metal to the graphene.

Since it is possible that Ni is reacting with EG to form end-contatcs, similar to the CVD graphene case,[15] we next determine the effect of metal deposition on graphene lattice integrity. Four 4x4 mm$^2$ epitaxial graphene samples (EG_1, 2, 3, 4; synthesized from one semi-insulating nominally on-axis SiC(0001) wafer) were patterned to contain large-area vdP Hall structures with Ti/Au (10 nm/300 nm) contacts deposited using e-beam evaporation through a shadow mask. Two of these samples (EG_2 and EG_4) were subsequently exposed to lift-off resist and S1811 (MICROPOSIT), flood exposed in deep UV and developed by CD-26 to simulate the effect of resist residue before metal deposition. 50 nm Ni was deposited by e-beam evaporation on all four EG samples and two CVD grown graphene samples (CVD_1, 2) after transferring on Si/SiO$_2$ by standard wet-transfer.[29] Samples (EG_3, EG_4 and CVD_2) were then annealed at 200°C under vacuum (~$10^{-4}$ mbar) for 24 hrs prior to etching the Ni film. The Ni was then etched by Ni etchant TFB at RT and confirmed by optical microscope and AFM. It was found that a much longer time (~15- 20 min.) was needed to completely etch the annealed Ni films as opposed to the un-annealed samples (~5 min.).

The Raman spectra of EG and CVD samples are depicted in Fig. 5(a) and Fig. 5(b), respectively. The EG samples did not exhibit any defect peak (D peak ~1350 cm$^{-1}$) due to Ni deposition before or after vacuum annealing. However, the CVD films exhibited an increase in the D peak with the introduction of Ni. Additionally, the CVD films with the combination of Ni deposition and anneal treatment, have a prominent D peak indicative of significant damage to the underlying graphene lattice, as shown in Fig. 5 (b). Such a finding is in congruence with previous reports. [15, 19] Unlike CVD samples, as Ni deposition on EG does not produce end-contact, we conclude that in the discussion on annealing (above) the improvement in $R_C$ is due to proximity to the graphene plane.

To gain more insight about the role of resist contamination at the graphene-metal interface, XPS was performed on all four samples with a Thermo Scientific k-Alpha system (spot size ~400 μm) using Al k-$\alpha$ radiation. Both the samples (EG_2 and EG_4) which were exposed to polymers showed (Fig. 5(c)) an additional peak at 288.4 eV which are attributed to either $NiCO_3$ or $Ni(CO)_4$.[30] However, samples with such contaminations showed similar or higher sheet resistance than those where no such peaks were observed (EG_1 and EG_3). Hence we infer that such contaminations are providing higher resistance paths than the graphene surface. Such non-conducting residues at graphene-metal interface potentially hinder carrier transmission which, incidentally, may alter the contact resistance.

Nevertheless, two important conclusions can be drawn from this study. First, a clean graphene-metal interface is not sufficient to obtain a $R_C$ limited only by quantum resistance, even though that might be a necessary condition for conformal metal deposition and reproducible contact resistance. Second, unlike CVD graphene, metal deposition does not form spontaneous end-contacts in epitaxial graphene, hence intentional edge-state conduction formation is required to obtain a low $R_C$.[15]

In summary, we demonstrate femto-second laser assisted direct patterning of graphene microstructures that enables us to study both intrinsic and extrinsic effects on the graphene-metal interface on the graphene planar surface without modifying the graphene by photoresist polymer residue or other chemicals. We show that a clean graphene-Ni interface does not lead to $R_C$ approaching the intrinsic limit set by the quantum resistance. We also found that the $R_C$ is primarily limited by graphene-Ni vertical carrier transmission and the effect of induced doping by the metal. Furthermore, we showed that the Ni interactions with EG when deposited by e-beam evaporation are much weaker than that of the CVD graphene, before or after annealing. Lastly, annealing probably is increasing the proximity of metal to the graphene plane which facilitates improved $R_C$. Yet, planar Ni contacts have $R_C$ over an order of magnitude higher than the calculated value, implying that edge state conduction is necessary to obtain $R_C$ near the quantum contact resistance limit.

**Acknowledgement.** M.J.T and Z.R.R are grateful for ASEE postdoctoral fellowship. We acknowledge useful discussions with M. S. Osofsky, D.J. Meyer (U.S. Naval Research Laboratory) and M.S.Fuhrer ( Monash University) . This work was sponsored by the Office of Naval Research.

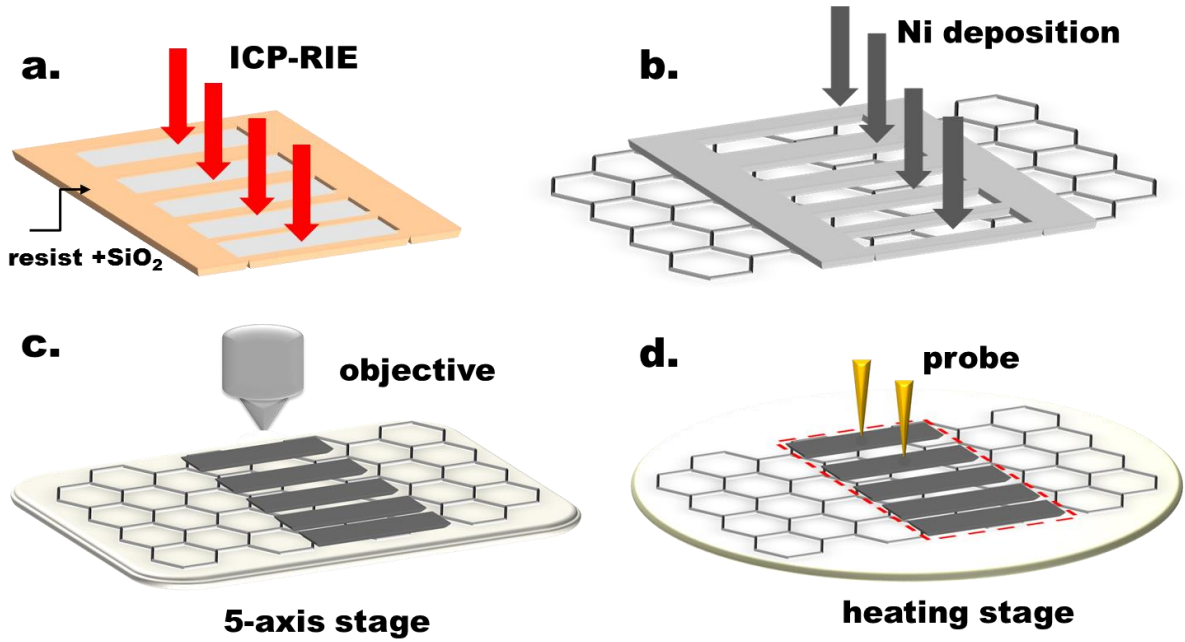

FIG.1. (Color online) Process steps: (a) Fluorine assisted ICP-RIE to prepare Si shadow mask. A stack of photoresist and SiO$_2$ was used as etch mask . (b) 50 nm Ni deposited by e-beam evaporator. (c) Device isolation by femtosecond laser. (d) *In-situ* annealing and electrical measurements. The red box shows isolated devices.

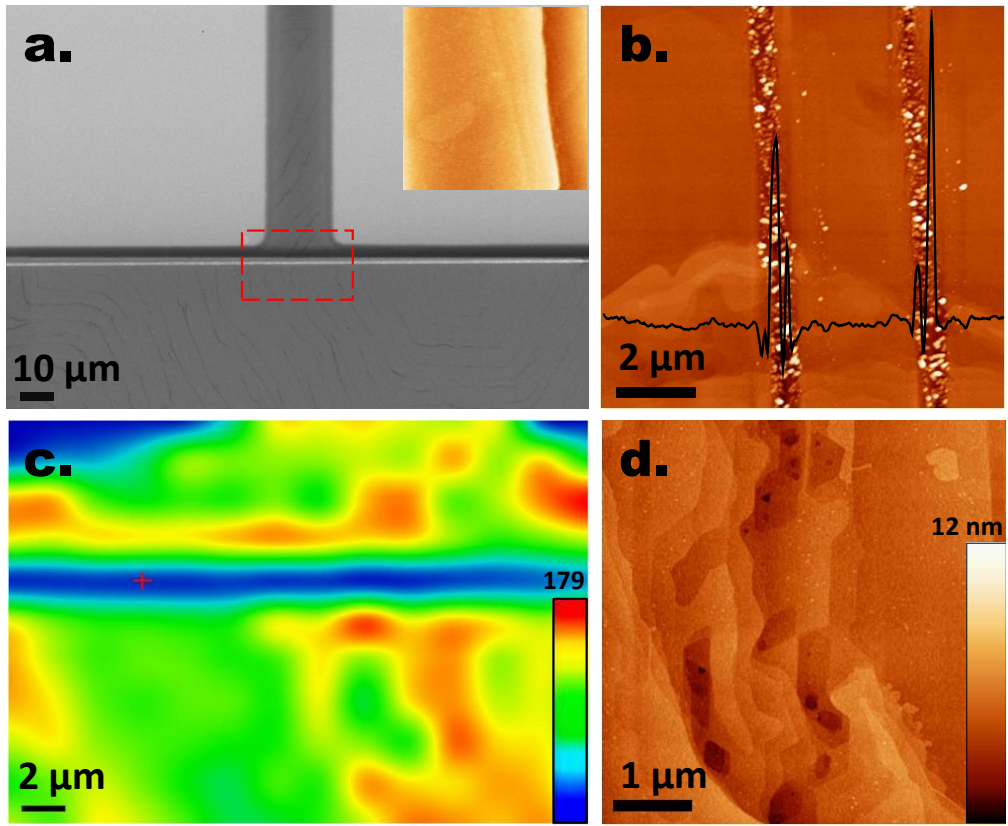

FIG.2. (Color online) InLense SEM image of an isolated TLM structure. The white straight line is the laser ablation. Inset 3 × 3 µm AFM height of image of Ni pad on graphene (rms roughness on terrace=0.24 nm). (b) AFM height image of isolation between adjacent TLM structures. (c) Raman 2D intensity map of the boxed area shown in (a). Color bar: arbitrary units. (d) AFM height image of graphene surface 5 µm away from the ablated lines shown in (b).

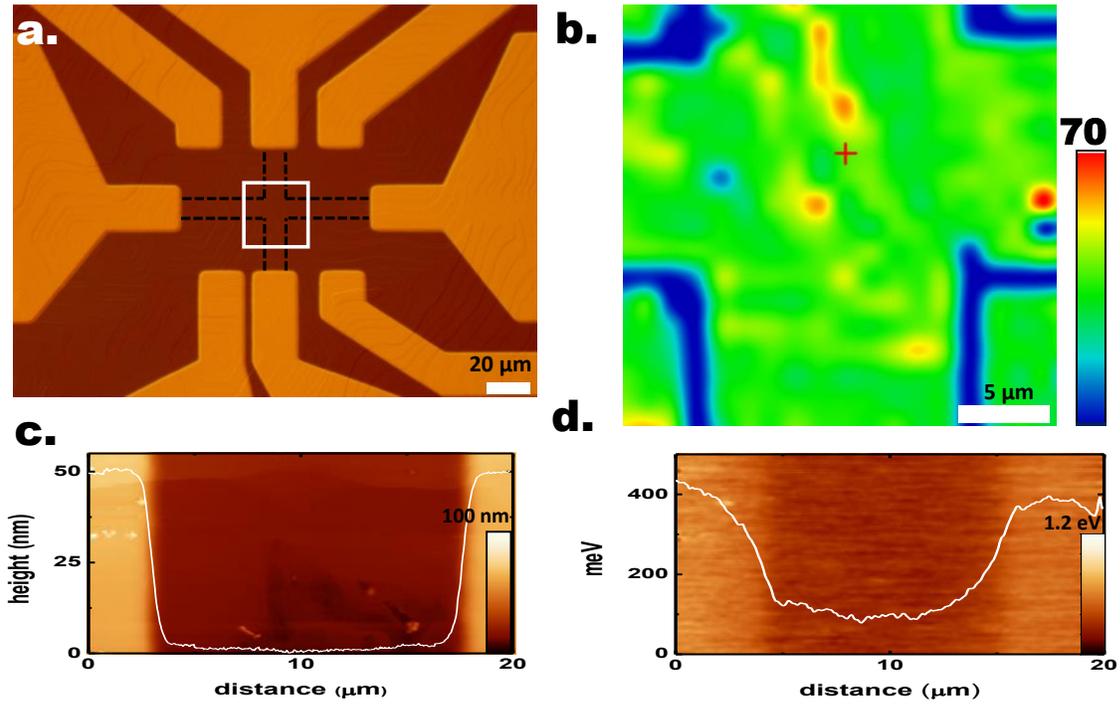

FIG.3. (Color online) Optical DIC image of a representative Hall structure fabricated by shadow mask. (b) Raman 2D FWHM map of center portion of an isolated van der Pauw cross structure as shown in (a). (c) AFM and (d) KPFM height images of an isolated TLM pads.

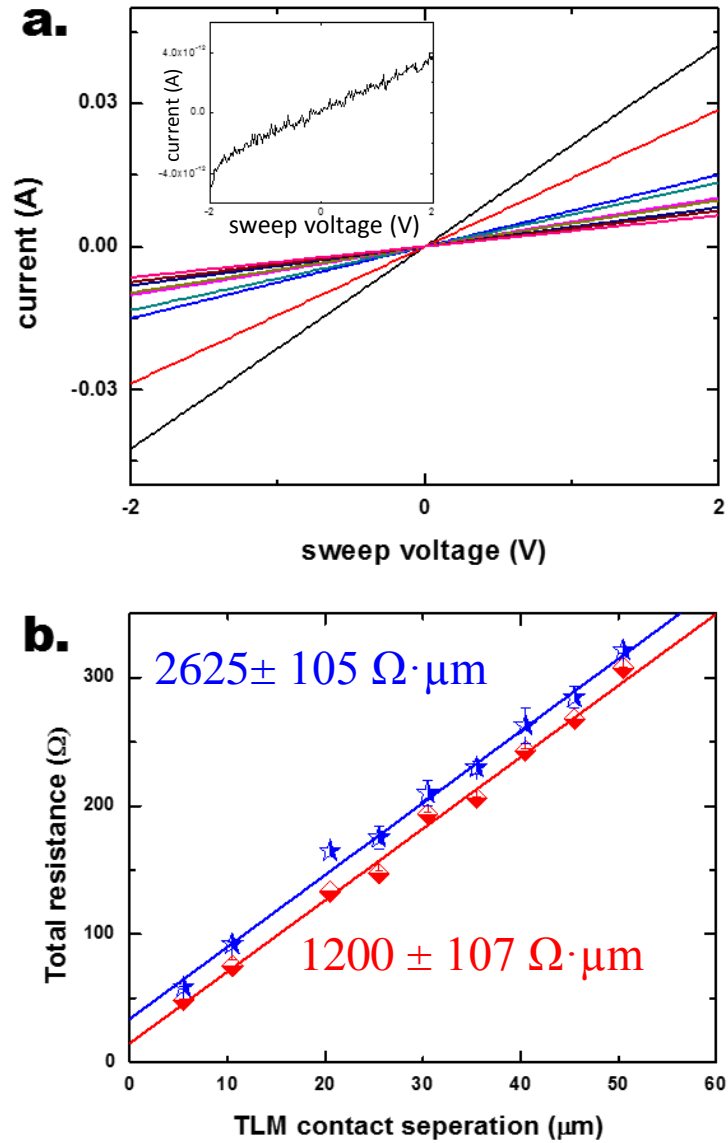

FIG.4. (Color online) RT I-V curves for different TLM separations after *in-situ* anneal showing Ohmic behavior. Inset shows isolation current. (b) TLM results before (blue stars) and after (red squares) *in-situ* annealing.

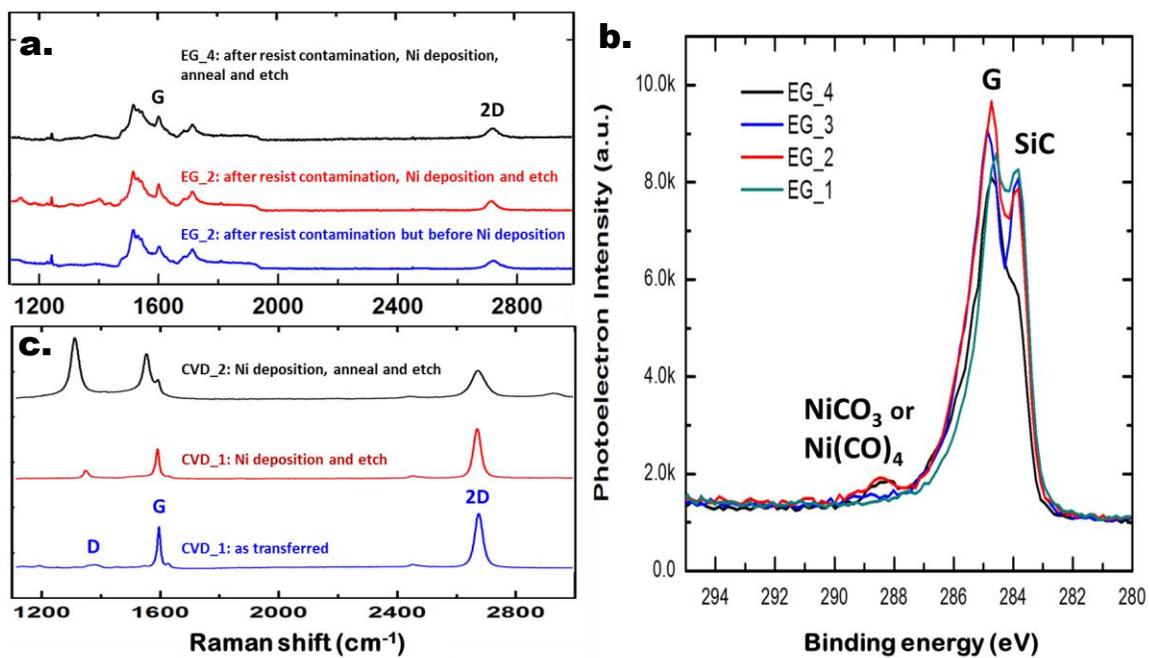

FIG.5. (Color online) Raman spectroscopy of graphene grown on SiC before and after Ni deposition and anneal. (b) XPS data of four different samples as described in the text after Ni deposition and etch or Ni deposition, annealed and etched. (c) Raman spectroscopy of CVD grown samples before and after Ni deposition and anneal.